\documentclass[prx,twocolumn,showpacs,superscriptaddress,dvipsnames]{revtex4-1}

\usepackage[SquareTraceBrackets]{quantum}
\usepackage{graphicx,xcolor,upgreek,amsbsy,amsmath,eufrak,array}

\usepackage[colorlinks=true,citecolor=MidnightBlue,linkcolor=MidnightBlue,urlcolor=MidnightBlue]{hyperref}

\usepackage{tgtermes}
\usepackage[T1]{fontenc}
\usepackage[scaled]{helvet}

\newcommand{\bmu}{\boldsymbol{\upmu}}

\newcommand{\btheta}{\boldsymbol{\theta}}
\newcommand{\dif}[2]{\, \mathrm{d}^{#2} #1 \,}

\def\NO#1{\mathop{:}\nolimits\!#1\!\mathop{:}\nolimits}
\DeclareMathOperator{\cov}{Cov}
\DeclareMathOperator{\var}{Var}
\DeclareMathOperator{\perm}{Perm}

\begin{document}

\title{\large\textsf{\textbf{Precision Estimation of Source Dimensions from Higher-Order Intensity Correlations}}}

\author{M. E. Pearce}\email{mrkprc1@gmail.com}
\affiliation{Department of Physics \& Astronomy, University of Sheffield, Sheffield S3 7RH, United Kingdom}

\author{T. Mehringer}
\affiliation{Institut f\"ur Optik, Information und Photonik, Universit\"at Erlangen-N\"urnberg, 91058 Erlangen, Germany}
\affiliation{Erlangen Graduate School in Advanced Optical Technologies (SAOT)}

\author{J. von Zanthier}
\affiliation{Institut f\"ur Optik, Information und Photonik, Universit\"at Erlangen-N\"urnberg, 91058 Erlangen, Germany}
\affiliation{Erlangen Graduate School in Advanced Optical Technologies (SAOT)}

\author{P. Kok}\email{p.kok@sheffield.ac.uk}
\affiliation{Department of Physics \& Astronomy, University of Sheffield, Sheffield S3 7RH, United Kingdom}

\begin{abstract}
 \noindent 
An important topic of interest in imaging is the construction of protocols that are not diffraction limited. This can be achieved in a variety of ways, including classical superresolution techniques or quantum entanglement-based protocols. Here, we consider superresolving imaging in the far field using higher-order intensity correlations. We show that third and fourth order correlations can improve upon the first and second order correlations that are traditionally used in classical optics and Hanbury Brown and Twiss type experiments. The improvement is achieved entirely by post-processing of the data. As a demonstrator, we simulate the far field intensity distribution of a circular aperture that emits thermal light and use maximum likelihood estimation to determine the radius of the aperture. We compare the achieved precision to the Cram\'er-Rao lower bound and find that the variance of measurements for the third and fourth order correlation functions are indeed closer to the Cram\'er-Rao bound than that of the second order correlation function. The method presented here is general, and can be used for all kinds of incoherent emitters, geometries, and types of noise.
\end{abstract}

\date{\today}

\maketitle  

\section{introduction}\noindent
Visualization of natural phenomena has played a central role in the development of science and technology. Indeed, the invention of the telescope shed light on the motion of planets and stars and the conception of the microscope allowed investigations into the ingredients of life at the microscopic level. Historically, every time a new imaging technique was introduced, science has leaped a step forward. Most recent advances in imaging include exoplanet detection \cite{Macintosh14} and the velocity measurement of molecular markers along DNA \cite{wuite13}. However, the wave nature of light dictates that there are physical limits to the resolution of optical telescopes and microscopes, as formulated by Abbe in 1873 \cite{Abbe1873}. In order to see smaller details in microscopy we could illuminate with light of shorter and shorter wavelengths, but this is not always practical: highly energetic light may destroy biological samples, and in astronomy shorter wavelengths are increasingly difficult to access. It is thus useful to find alternative techniques to improve the resolution of imaging methods that overcome the diffraction limit. Imaging techniques that yield finer details than that dictated by the Abbe limit are referred to as \emph{superresolving} techniques.  In microscopy techniques such as photo-activated localised microscopy ({\sc palm}) \cite{Hess2006}, stochastic optical reconstruction microscopy ({\sc storm}) \cite{Rust2006} and stimulated-emission depletion microscopy ({\sc sted}) \cite{Hell1994} achieve superresolved images using fluorescent markers. These methods combine standard intensity measurements with non-linear effects or stochastical processes in combination with prior information about the sample preparation and post-processing to achieve the pursued goal of superresolution. 

\begin{figure}[b]
\centering
\includegraphics[width=8.5cm]{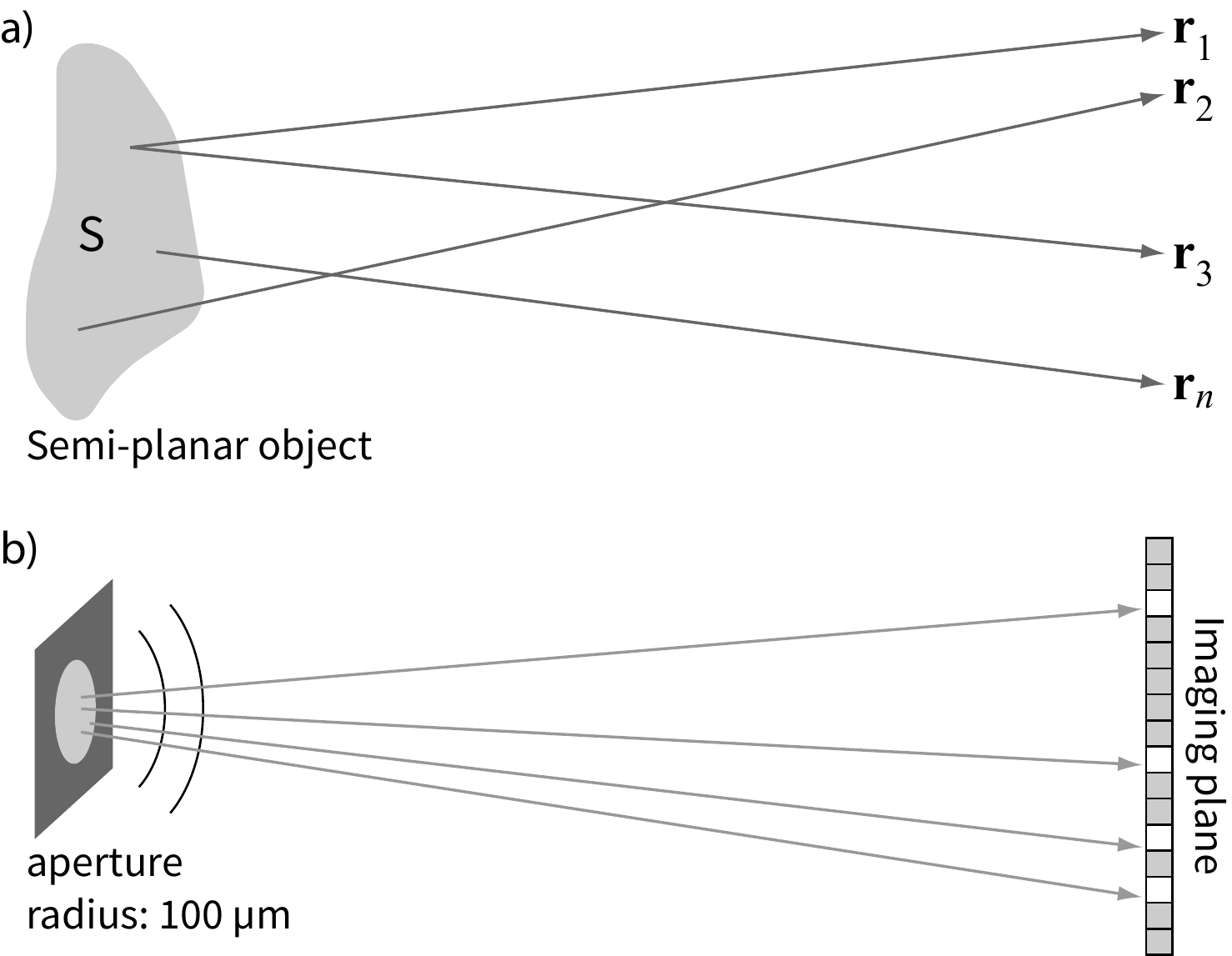}
\caption{(a) Spatially incoherent, planar source $S$ emitting into the far field.  The intensity measured at the positions ${\bf r}_1, \dots, {\bf r}_n$ can be correlated to give the $n$-point intensity correlation function $G^{(n)}(\textbf{r}_1, \ldots, \textbf{r}_n)$, which contains information about the geometry of $S$. (b) Implementation simulated here: a circular aperture of radius $100~\upmu$m emits monochromatic thermal light with a wavelength of 633~nm, which is recorded with a CCD camera. The intensity measured by the different pixels of the CCD can be correlated to calculate $G^{(n)}(\textbf{r}_1, \ldots, \textbf{r}_n)$ as in (a). The task is to estimate the radius of the aperture from $G^{(n)}(\textbf{r}_1, \ldots, \textbf{r}_n)$.}\label{fig:geo}
\end{figure}

Alternatively, superresolving imaging may also be achieved using higher-order intensity interferometry \cite{Thiel07,Oppel12,Monticone2014}. Hanbury Brown and Twiss (HBT) in their seminal experiments demonstrated that the second order intensity correlation function is proportional to the Fourier transform of the intensity distribution of a thermal light source \cite{BrownTwiss56}.  Measuring the intensity-intensity correlations therefore provides a method to access the spatial distribution of light emitters. The question then arises how and to what extent higher-order intensity correlations can be used to improve the resolution in imaging. Early proposals using entangled $N$-photon states promised an increase in resolution $\propto O(N)$ \cite{Boto2000}. Recent more practical multi-photon correlation techniques achieve a similar scaling \cite{Thiel07,Oppel12,Monticone2014}, relying on the precise estimation of certain parameters.  To ensure the optimal performance of such techniques it is thus necessary to obtain the best estimates possible. In this paper for the first time rigorous estimation theory is used to demonstrate how higher-order intensity correlation measurements can yield a resolution improvement over conventional second order intensity interferometry. The general setup is shown in Fig.~\ref{fig:geo}. As a demonstrator we investigate a planar circular source that emits spatially incoherent monochromatic thermal light, which is recorded in the far field by $n$ detectors at $\textbf{r}_1, \ldots, \textbf{r}_n$. The recorded intensities are then correlated in order to produce the $n^{\rm th}$-order intensity correlation function. In this paper, we provide a general framework that allows us to investigate intensity correlations of arbitrary order and discuss the advantages and disadvantages with respect to improved resolution produced by higher-order intensity correlation measurements.  We present a rigorous analysis of the HBT set up, explicitly calculating the corresponding likelihood function and Fisher information, allowing us to determine the lowest possible variance in estimates of the source. We also perform a maximum likelihood estimation, enabling us to achieve this bound.  An integral part of our procedure is to treat unknown quantities as parameters to be estimated, every unknown therefore increasing the size of the estimation problem.  We then demonstrate the method with a concrete example, making use of simulated data, to obtain precision estimates of the radius of a thermal source.  We also show how these results can be used to estimate the dimensions of any source, regardless of its geometry or its photon statistics, as long as the total photon number is not deterministic (i.e. $\Delta \hat{n} > 0$). For any such source, our method allows us to calculate the Fisher information, providing maximum likelihood estimates of the source dimensions via the method of scoring.  By explicitly calculating the Fisher information for different correlation orders $n$, we then show that in some instances the Fisher information increases with correlation order. The exact relationship between the Fisher information and the correlation order depends strongly on the particular arrangement of the detectors and the source geometry.  

The idea of exploiting spatial intensity correlations of order $n > 2$ to improve resolution has been suggested previously \cite{Thiel07,Oppel12,Iskhakov2011}. In particular in the context of ghost imaging \cite{Pittman95,Strekalov95,Lugiato2002,Shih07,Erkmen10}, the idea of exploiting higher correlation orders has received great attention \cite{Zhou2010,Chen2010,Chan2009,Agafonov2009,Chen10}. However, to date there has not been a rigorous theoretical approach to quantify the possible enhancement in resolution when using higher-order correlation measurements for HBT-type experiments.  We provide that explanation here and demonstrate its use.  The effects of photon losses are included in our model and we show that higher-order correlations can continue to perform well, even in the presence of loss.

The paper is organized as follows: In Section II we present the theoretical model used to determine the higher-order correlations of the intensity produced in the far field by the investigated source. We present the exact mathematical expression for the corresponding $n^{\rm th}$-order intensity correlation function and show that in the case of a thermal source it can be written in terms of the permanent of the correlation matrix involving the complex degree of coherence. In Section III we present the theory of parameter estimation, which involves the use of optimized estimators that take the measured data to return optimal estimates of the parameters of interest. This allows us to evaluate their performance by use of the Cram\'er-Rao bound, expressed in terms of the Fisher information, i.e., it enables us to derive an explicit expression for the lower bound of the variance in the estimates of the considered source. In Section IV we present an explicit protocol for obtaining the data of an $n^{\rm th}$-order intensity correlation measurement, incorporating statistical uncertainties as well as additional noise due to, e.g., the limited detection efficiency of the detectors. This enables us to calculate the covariance matrix for the given problem, which in turn gives access to the Fisher information, the Cram\'er-Rao bound, and the maximum likelihood estimation procedure. In Section V we present detailed numerical simulations for different source geometries and calculate the Cram\'er-Rao bound for the source dimension for different orders of the intensity correlation function, taking into account two kinds of noise, a constant noise factor at each detector and Gaussian distributed noise due to detector losses. Finally, in Section VI we discuss our results and present our conclusions.

\section{n-point intensity correlation functions}\noindent
We consider a semi-planar source $S$ emitting monochromatic spatially incoherent thermal radiation that is observed in the far field.  The setup is shown in Fig.~\ref{fig:geo}a. The second order intensity interference for experiments that measure the equal-time two-point intensity correlation reads
\begin{align}\label{eq:Gn}
G^{(2)}(\mathbf{r}_1, \mathbf{r}_2) & = \braket{ \NO{ \hat{E}^{(-)}(\mathbf{r}_1) \hat{E}^{(+)}(\mathbf{r}_1) \hat{E}^{(-)}(\mathbf{r}_2) \hat{E}^{(+)}(\mathbf{r}_2) }} \nonumber \\
& \propto  \braket{ \NO{ \hat{a}^{\dagger}(\mathbf{r}_1) \hat{a}(\mathbf{r}_1) \hat{a}^{\dagger}(\mathbf{r}_2) \hat{a}(\mathbf{r}_2) }}\, ,
\end{align}
where $E^{(\pm)}$ are the positive and negative frequency parts of the electric field, $\hat{a}(\mathbf{r})$ and $\hat{a}^{\dagger}(\mathbf{r})$ are the annihilation and creation operators of the field at position $\mathbf{r}$, $: \, :$ denotes normal ordering, and $\mathbf{r}_i$ is the position of the $i^{\mathrm{th}}$ detector in the far field.  When viewed in the far field of the source, interference fringes can be observed in $G^{(2)}({\bf r}_1,{\bf r}_2)$ under certain conditions \cite{BrownTwiss56}. However, we can also consider the equal-time $n$-point intensity correlation $G^{(n)} \propto \braket{\NO{ \prod_{i=1 }^n \hat{a}^{\dagger}(\mathbf{r}_i) \hat{a}(\mathbf{r}_i)}}$, as  increasing the correlation order $n$ can lead to an increased visibility of the interference fringes, suggesting we may be able to extract more information from the higher correlation orders \cite{Oppel12, Agafonov08}.    

\begin{figure}
\centering
\includegraphics[width=8.5cm]{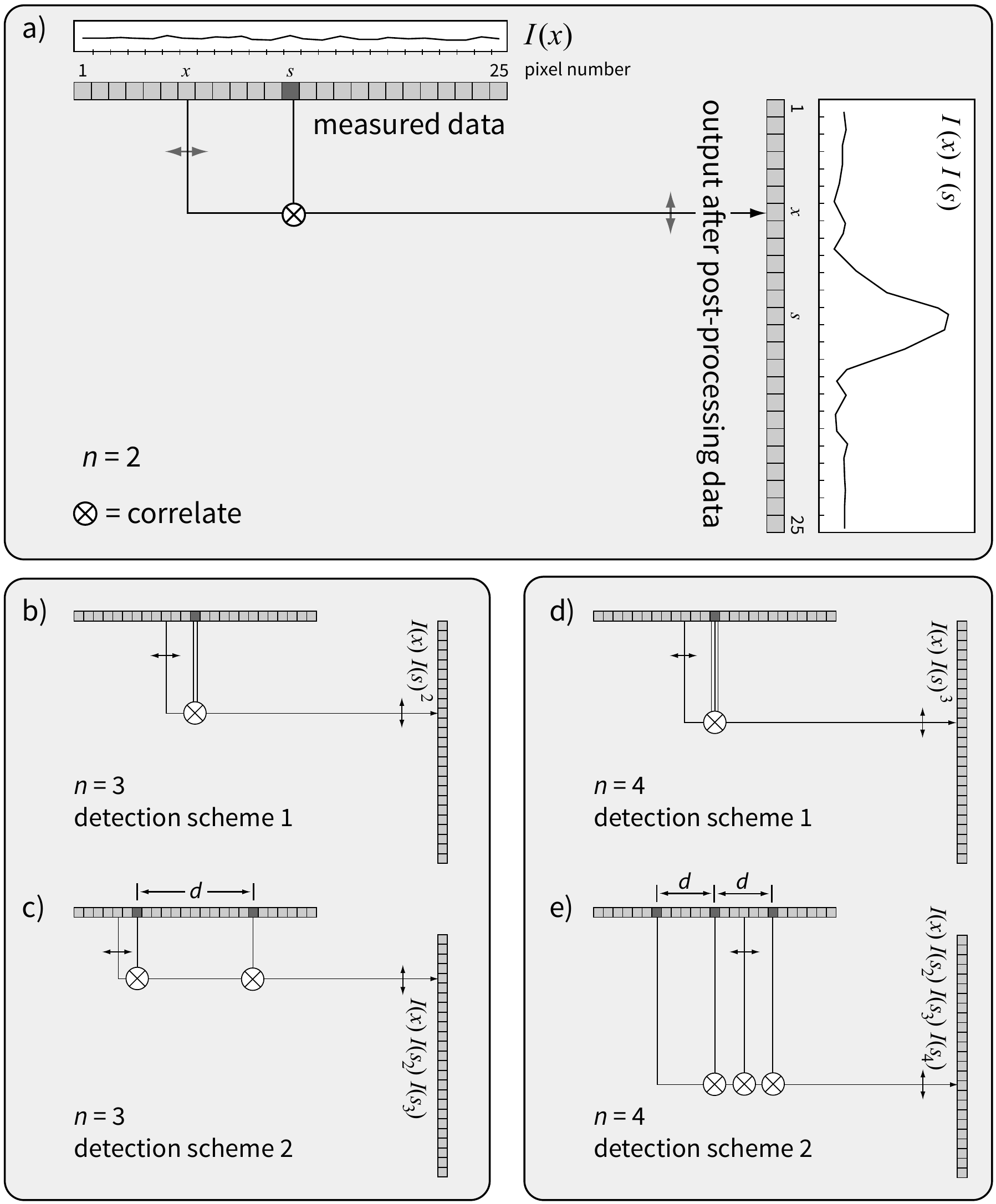}
\caption{Measuring the higher-order intensity correlation functions with an array of pixels.  (a) The second order intensity correlation at pixel $x_i = 7$ is calculated as the correlation between the intensities at a fixed pixel $s_2 = 13$ (shown as a darker pixel) and at the pixel $x_i = 7$. (b, c) The third order correlation is defined using two detection schemes, namely via a single fixed pixel that is correlated twice (detection scheme 1), or two fixed pixels (detection scheme 2). (d, e) The fourth order intensity correlation is defined analogous to the third order. The reference pixel separation $d$ is a dimensionless number, and the center-to-center separation of adjacent pixels is taken as $5.3~\upmu$m throughout the paper.}
\label{fig:pixarray}
\end{figure}

To obtain the higher-order intensity correlations we consider the particular measurement scheme displayed in Fig.~\ref{fig:pixarray}. An array of pixels arranged along an axis parallel to the surface of the source allows to measure the intensity at a discrete number of positions.  The benefit of such a setup is that we can capture a large amount of data in one frame.  In this section we establish the theoretical model of the intensity correlations to any order; the estimation procedure will be discussed in detail in Sections III and IV.

To calculate the $n$-point intensity correlation we make use of the optical equivalence theorem \cite{Sudarshan1963}, which states that the expectation of a normally ordered product of creation and annihilation operators can be replaced by their left and right eigenvalues, respectively, if the expectation is replaced by an ensemble average weighted by the $P$-representation of the state. That is, mathematically, we can write
\begin{align}
\braket{f(\hat{a}^{\dagger},\hat{a})} & = \int P(\alpha) f(\alpha^*,\alpha) \dif{\alpha}{2} \equiv \braket{f(\alpha^*,\alpha)}_{P},
\end{align}
where $f$ is any normally ordered function of the creation and annihilation operators, the first expectation is the quantum mechanical average, and the subscript $P$ on the second expectation signifies that it is an ensemble average taken with respect to the quasi-probability distribution $P$.  With the help of the optical equivalence theorem, the $n$-point intensity correlation $G^{(n)}(\mathbf{r}_1,\dots,\mathbf{r}_n)$ can thus be written as
\begin{align}
\Braket{ \NO{ \textstyle{\prod_{i=1}^n} \hat{a}^{\dagger}(\mathbf{r}_i) \hat{a}(\mathbf{r}_i) } } = \Braket{ \textstyle{\prod_{i=1}^n} \alpha^*(\mathbf{r}_i) \alpha(\mathbf{r}_i) }_{P}.
\end{align}
Thermal light exhibits a Gaussian zero mean $P$-representation.  We can therefore apply the Gaussian moment theorem \cite{MandelandWolf1995} and make the simplification
\begin{align}
\Braket{ {\textstyle{\prod_{i=1}^n}} \alpha^*(\mathbf{r}_i) \alpha(\mathbf{r}_i) }_{P} = \sum_{\sigma \in S_n} \prod_{i=1}^n \braket{\alpha^*(\mathbf{r}_i) \alpha(\mathbf{r}_{\sigma(i)})},
\end{align}
where $S_n$ is the symmetric group containing all permutations of the set $\{ 1, \dots ,n \}$.  This allows us to write $G^{(n)}$ as
\begin{align}
G^{(n)}(\mathbf{r}_1,\dots,\mathbf{r}_n)  = |K|^{2n} \sum_{\sigma \in S_n} \prod_{i=1}^n \braket{ \hat{a}^{\dagger}(\mathbf{r}_i) \hat{a}(\mathbf{r}_{\sigma(i)})  } ,
\end{align}
where we have defined $\hat{E}^{(+)} = K \hat{a}$.  We can make a further simplification by introducing the complex degree of coherence, defined as \cite{MandelandWolf1995}\footnote{the complex degree of coherence is sometimes called the \emph{mutual coherence function} \cite{GerryandKnight}}
\begin{align}
\gamma(\mathbf{r}_1, \mathbf{r}_2) = \frac{\braket{ \hat{a}^{\dagger}(\mathbf{r}_1) \hat{a}(\mathbf{r}_2)}}{\left[ \braket{ \hat{a}^{\dagger}(\mathbf{r}_1) \hat{a}(\mathbf{r}_1)} \braket{ \hat{a}^{\dagger}(\mathbf{r}_2) \hat{a}(\mathbf{r}_2)} \right]^{1/2} } .
\end{align}
This allows us to write the $n$-point intensity correlation function as
\begin{align}\label{eq:Gntherm}
G^{(n)}(\mathbf{r}_1,\dots,\mathbf{r}_n) & = \sum_{\sigma \in S_n} \prod_{i=1}^n \braket{\hat{a}^{\dagger}(\mathbf{r}_i) \hat{a}(\mathbf{r}_i)} \gamma(\mathbf{r}_i , \mathbf{r}_{\sigma(i)})\, ,
\end{align}
where we omitted the constant of proportionality $K$ for brevity.  From this expression we see that the $n$-point correlation for Gaussian light is equal to the permanent of a matrix $\Gamma$
\begin{align}
G^{(n)}(\mathbf{r}_1,\dots,\mathbf{r}_n) & = \perm (\Gamma)\, , 
\end{align}
where
\begin{align}
\Gamma_{ij} & \equiv \left[ \braket{ \hat{a}^{\dagger}(\mathbf{r}_i) \hat{a}(\mathbf{r}_i)} \braket{ \hat{a}^{\dagger}(\mathbf{r}_j) \hat{a}(\mathbf{r}_j)} \right]^{1/2} \gamma(\mathbf{r}_i, \mathbf{r}_j).
\end{align}
In general the permanent of a matrix is difficult to calculate.  Therefore, for larger $n$ it is increasingly costly to calculate the $n$-point correlations, this may be the main limiting factor for the use of higher-order correlation functions in imaging.

The advantage of writing $G^{(n)}(\mathbf{r}_1,\dots,\mathbf{r}_n)$ in terms of the complex degree of coherence is that the complex degree of coherence in the far field paraxial regime is given by the two-dimensional Fourier transform of the intensity distribution of the source \cite{MandelandWolf1995}. This result is known as the Van Cittert-Zernike theorem \cite{VanCittert34,Zernike1938}.   We can therefore calculate the far field intensity correlations of any order for any source geometry, provided that we can determine the Fourier transform of the intensity distribution.

\begin{figure}
\centering
\includegraphics[width=8.5cm]{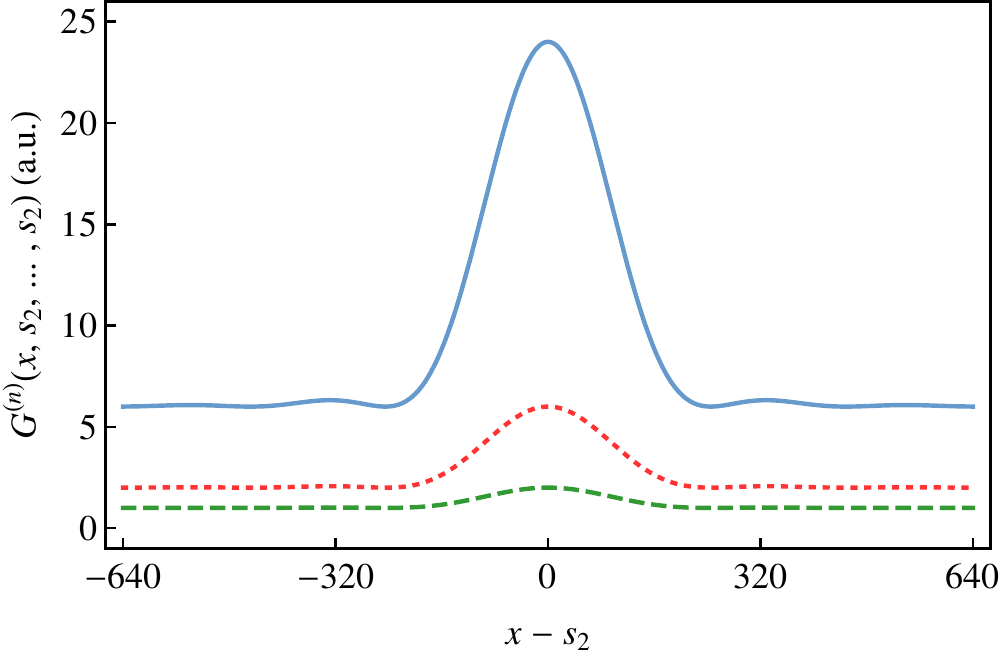}
\caption{(Color online) The second (\textcolor{OliveGreen}{dashed}), third (\textcolor{red}{dotted}) and fourth (\textcolor{MidnightBlue}{solid}) order intensity correlation functions of Eq.~(\ref{eq:corf}) as a function of the separation between the scanning pixel $\mathbf{r}_1 = x$ and the reference pixels at $s_2$ for the remaining arguments, with normalised intensity $\braket{\hat{a}^{\dagger}(\mathbf{r}_i) \hat{a}(\mathbf{r}_i)} = 1$ and $\vartheta = 5\times 10^{-4}$~rad. The width of the curves is directly proportional to the angular diameter of the source $\vartheta$. The higher visibilities of \textcolor{red}{$G^{(3)}$} and \textcolor{MidnightBlue}{$G^{(4)}$} over \textcolor{OliveGreen}{$G^{(2)}$} suggest that higher-order correlation functions may outperform regular Hanbury Brown and Twiss estimation of the source diameter. In this paper, we show rigorously using estimation theory that this can indeed be the case and specify the conditions.}
\label{fig:G2plot}
\end{figure}

As an example, consider a circular source of uniform intensity and angular diameter $\vartheta = 2 \tan^{-1} (a/L) \approx 2a/L$ where $a$ is the radius of the source and $L$ is the distance from the source to the observation plane.  The far field complex degree of coherence is the two-dimensional Fourier transform of a circle with radius $a$ \cite{BornandWolf}
\begin{align}\label{eq:cdccirc}
\gamma(\mathbf{r}_1 , \mathbf{r}_{2})  = \frac{2 J_1 \left( \frac{1}{2}\vartheta k | \mathbf{r}_{1} - \mathbf{r}_{2} | \right)}{\left( \frac{1}{2}\vartheta k | \mathbf{r}_{1} - \mathbf{r}_{2} | \right)}, 
\end{align}
where $J_1$ is the first order Bessel function of the first kind and $k$ is the wavenumber.  The $n$-point correlation function then becomes
\begin{align}\label{eq:corf}
G^{(n)}(\mathbf{r}_1,\dots,\mathbf{r}_n) & = \nonumber \\
 \sum_{\sigma \in S_n} \prod_{i=1}^n & \braket{\hat{a}^{\dagger}(\mathbf{r}_i) \hat{a}(\mathbf{r}_i)} \frac{2 J_1 \left( \frac{1}{2} \vartheta k | \mathbf{r}_{i} - \mathbf{r}_{\sigma(i)} | \right)}{\left( \frac{1}{2} \vartheta k | \mathbf{r}_{i} - \mathbf{r}_{\sigma(i)} | \right)}.
\end{align}
Fig.~\ref{fig:G2plot} shows the second, third and fourth order intensity correlation functions for a uniform disc along a one-dimensional detection device.

\section{Estimation theory}\noindent
In the previous section we saw that intensity correlation measurements in the far field depend on the parameters describing the source geometry. In this section we examine how we can use these measurements to practically obtain spatial information about the source.  To this end we employ parameter estimation theory, which involves the use of an estimator $\hat{\btheta}$ that takes the measured data ${\bf x} = (x_1,\ldots,x_M)$ in $M$ measurements (the pixels in the detector) and returns estimates of the parameters of interest $\btheta = (\theta_1,\dots, \theta_l)$.  In order to extract the spatial information in the most efficient way, we can apply a maximum likelihood estimation procedure.  Maximum likelihood estimation relies on maximisation of the joint probability distribution of our data, and we therefore need to characterise the probability distribution from which the correlation functions are sampled \cite{Kay1993}.   

Once we have determined the conditional probability $p({\bf x }| \btheta)$ of obtaining the measurement outcomes ${\bf x}$ given the values of the parameters $\btheta = (\theta_1, \dots , \theta_l)$, we can determine the performance of our estimates using the Cram\'er-Rao bound (CRB).  The CRB provides a lower bound for the variance of our estimators in terms of the Fisher information matrix
\begin{align}
\var(\theta_i) \geq [{\bf I}(\btheta)]_{ii}^{-1},
\end{align}
where the Fisher information matrix ${\bf I} (\btheta)$ is given by 
\begin{align}\label{eq:FIM}
[ \mathbf{I}(\btheta)]_{ij} = \sum_{\mathbf{x}} p(\mathbf{x} | \btheta) \left( \frac{\partial \ln[p(\mathbf{x}|\btheta)] }{\partial \theta_i} \right) \left( \frac{\partial \ln[p(\mathbf{x}|\btheta)] }{\partial \theta_j} \right) .
\end{align}
Note that the sum over $\mathbf{x}$ may be an integral if the values of $x_i$ form a continuum. In practice, intensity measurements in modern optical detectors yield a digital signal with discrete values.

The exact probability distribution $p(\mathbf{x}|\theta)$ of the correlation functions will be a complicated expression depending on the number of images $N$---not to be confused with the number of pixels $M$ in each image.  However, since the measurements of the correlation functions are averages over a (preferably) large data set, the central limit theorem dictates that these measurements will be normally distributed \cite{vdvaart1998}.  Assuming that we make $N$ measurements of the correlation functions at $M$ discrete detector positions, the data will follow an $M$-dimensional normal distribution:
\begin{align}\label{eq:loglike}
p(\mathbf{x} | \btheta) = \frac{\exp \left(-\frac{1}{2}(\mathbf{x} - \bmu(\btheta))^{\rm T} \mathbf{C}^{-1}(\btheta) (\mathbf{x} - \bmu(\btheta)) \right) }{\sqrt{(2 \pi )^M | \mathbf{C}(\btheta) |}}\, ,
\end{align}
where $\bmu$ is the tuple of expectation values of the distribution at each of the sampling points: $\bmu^{\rm T} = (\braket{{\bf x}_1}, \dots, \braket{{\bf x}_M})$, $^{\rm T}$ denotes the transpose, and $\mathbf{C}$ is the covariance matrix between pairs of measurements $\mathbf{C}_{ij} = \cov({\bf x}_i,{\bf x}_j) = \braket{{\bf x}_i {\bf x}_j} - \braket{{\bf x}_i} \braket{{\bf x}_j}$.  

For a multivariate-normal distribution the elements of the Fisher information matrix are given by \cite{Kay1993}
\begin{align}
\label{eq:Fisher1}
[\mathbf{I}(\btheta)]_{ij} & = \left( \frac{\partial \bmu}{\partial \theta_i} \right)^{\mathrm{T}} \mathbf{C}^{-1} \left( \frac{\partial \bmu}{\partial \theta_j} \right) + \frac{1}{2} \Tr{ \mathbf{C}^{-1}\frac{\partial \mathbf{C}}{\partial \theta_i}\mathbf{C}^{-1}\frac{\partial \mathbf{C}}{\partial \theta_j}} \nonumber \\
& \equiv [\mathbf{I}_1(\btheta)]_{ij} + [\mathbf{I}_2(\btheta)]_{ij},
\end{align}
where we define the first term (depending on $\bmu$) as $[\mathbf{I}_1(\btheta)]_{ij}$, and the second term as $[\mathbf{I}_2(\btheta)]_{ij}$.  Since the optical field exhibits strong transverse correlations in the detection plane, the covariances will not be negligible, and we must therefore explicitly evaluate these covariances in order to perform maximum likelihood estimation. In the next section we explain how the correlation functions are measured and incorporated into the parameter estimation procedure.

Finally, in order to find the maximum likelihood estimate we use an iterative method called \emph{scoring} \cite{Kay1993}. The process is described by the recursion relation 
\begin{align}
\mathbf{I}(\btheta^{(k)}) \btheta^{(k+1)} = \mathbf{I}(\btheta^{(k)}) \btheta^{(k)} + \frac{\partial \ln [p(\mathbf{x}|\btheta)]}{\partial \btheta} \bigg|_{\btheta = \btheta^{(k)}},
\end{align}
where $\btheta^{(k)}$ is the $k^{\rm th }$ iteration of the parameters $\btheta$, and $X |_{\btheta = \btheta^{(k)}} $ denotes evaluation of the quantity $X$ at the value $\btheta = \btheta^{(k)}$.  In order to begin the scoring algorithm we require an initial value $\btheta^{(0)}$.  Provided the initial value is sufficiently close to the actual value, the algorithm should continue without difficulty.  If no prior knowledge exists about the parameters to be estimated, approximate values can be obtained by simple methods (which by no means achieve the CRB) that can then be used as the initial values $\btheta^{(0)}$.

\begin{table*}[t]
\begin{ruledtabular}
\begin{tabular}{lll}
  & statistics & $n^{\rm th}$-order intensity correlation measurements \\[5pt] 
 random variable & $X_i$  & $I(x_i)\, I(s_2) \ldots\,  I(s_n)$ \\[2pt] 
 $k^{\rm th}$ measurement & $X_i^{(k)}$  & $I_k(x_i)\, I_k(s_2) \ldots\,  I_k(s_n)$  \\[2pt] 
 sample mean & $\overline{X}_i = \frac{1}{N} \sum_{k=1}^N X_i^{(k)}$ & $\mathfrak{G}^{(n)}(x_i) = \frac{1}{N} \sum_{k=1}^N I_k(x_i) I_k(s_2) \ldots I_k(s_n)$  \\[2pt] 
 first moment & $\braket{\overline{X}_i} = \braket{X_i}$ & $\braket{\mathfrak{G}^{(n)}(x_i)} = \braket{I(x_i)\, I(s_2) \ldots\,  I(s_n)}  \propto G^{(n)}(x_i, s_2,\ldots,s_n)$ \\[2pt] 
covariance & $\cov \left( \overline{X}_i , \overline{X}_j \right) = \frac{1}{N} \cov \left(X_i , X_j \right)$ & $\cov \left(\mathfrak{G}^{(n)}(x_i) , \mathfrak{G}^{(n)}(x_j) \right) = \frac{1}{N} \cov \left(I(x_i) \, I(s_2) \ldots\,  I(s_n) , I(x_j) \, I(s_2) \ldots\,  I(s_n) \right)$
\end{tabular}
\end{ruledtabular}
\caption{Statistical quantities and their counterparts in the $n^{\rm th}$-order intensity correlation measurements. The random variable $X_i$ is the product of the intensity measurements at positions $x_i$, $s_2$,\ldots, $s_n$, and the index $i\in\{1,\ldots,M\}$ runs over all the detector positions (pixels). After $N$ measurements, we define a sample mean $\overline{X}_i$ that is itself a fluctuating quantity. This is not to be confused with the first moment $\braket{X_i}$, which is not a random variable. Since the sample mean is an unbiased estimate of the first moment, the expectation value of the sample mean is equal to the first moment. The covariance of any pair of sample averages is not equal to the covariance of the variables but $\smash{N^{-1}}$ times the covariance. This quantifies the fact that taking more data (increasing $N$) reduces the variation of the sample averages.  As $N \to \infty$ the sample averages coincide with the expectation values and the $\overline{X}_i$ are no longer random variables. The first moment is only proportional to the correlation function $G^{(n)}$  due to the efficiency factor in the measured intensity in Eq.~(\ref{eq:efficiency}).\label{tab:stats}}
\end{table*}

\section{Measuring the correlation functions}\noindent
For simplicity we suppress the $y$ dependence in ${\bf r}_i = (x_i,y_i)$ and consider only the one-dimensional problem where a single ``moving'' detector $x_i$ scans across a set of $M$ discrete positions $x_1,\dots,x_M$ and the remaining $n-1$ detectors are kept fixed (see Fig.~\ref{fig:pixarray}b-e).  We refer to the fixed detectors as the reference pixels and write the reference pixel positions as $x_2=s_2,\dots, x_n=s_n$. We distinguish between two detection schemes, namely one where all reference pixels are identical (scheme 1), and one where all reference pixels are different (scheme 2). Taking $N$ images means performing $N$ independent measurements of the intensity $I(x_i)$ at all pixels $x_i = x_1, \dots , x_M$ and calculating the sample average of the intensity moments
\begin{align}\label{eq:gnmeas}
\mathfrak{G}^{(n)}(x_1, \dots, x_n) & = \frac{1}{N} \sum_{k=1}^N \prod_{i=1}^n I_k(x_i)\, ,
\end{align}
where $I_k(x_i)$ is the $k^{\mathrm{th}}$ measurement of the intensity at position $x_i$.  Given the reference pixels $\{s_2,\ldots, s_n\}$, we can abbreviate $\mathfrak{G}^{(n)}(x_i,s_2,\dots,s_n) \equiv \mathfrak{G}^{(n)}(x_i)$, and the data that is used in the estimation procedure is then given by $\mathbf{x} = (\mathfrak{G}^{(n)}(x_1), \dots , \mathfrak{G}^{(n)}(x_M))$. This is still quite a general description and includes, for example, the experimental arrangement used in Ref.~\cite{Oppel12}. We use $\mathfrak{G}^{(n)}$ to denote a \emph{measurement} of the correlation function. This is not to be confused with the true correlation function $G^{(n)}$ as given by Eq.~\eqref{eq:Gntherm}.  It is an important distinction since the measured correlation $\mathfrak{G}^{(n)}(x_i)$ is a random variable due to the finite size of the sample $N$, whereas $G^{(n)}(x_i)$ is the expectation value of the correlation function, only in the limit $N\to\infty$ do the two coincide. The relation between the standard statistical quantities and the correlation functions are collated in Table \ref{tab:stats}.  
 
In addition to the statistical noise due to the finite sample size $N$, any realisable detection scheme will introduce additional noise into the measurements.  One important source of noise is reduced detection efficiency of the pixels. Often this is treated as a constant parameter $\eta$.  However, when calculating the intensity correlations we necessarily sample the higher moments of the detector noise.  It is therefore important that we acknowledge the random nature of the noise in order to correctly deduce its effects.  Physically we would expect the noise to be sharply peaked around some constant value with some small but non-zero variance.  We would also expect the random noise to be independent across the pixel array since the pixels themselves are independent.  We model this additional noise as uncorrelated, normally distributed noise with mean and variance $\braket{\eta(x_i)} = \nu$ and $\braket{\eta(x_i)^2} - \braket{\eta(x_i)}^2 = \varsigma^2$, respectively \footnote{The noise model can be chosen more generally: as we will show in the Appendix it is  necessary to determine only up to the $2n^{{\rm th}}$ moment of the noise distribution, treating each new moment as a new parameter to be estimated.  The use of Gaussian noise allows us to use only two parameters to calculate any moment, a feature not shared by a general distribution.  The additional parameters of a general noise model (higher moments) can be added to the list of estimation parameters, thereby increasing the size of the estimation problem whilst still keeping the problem tractable.}.  We can therefore write
\begin{align}\label{eq:efficiency}
I_k(x_i) =  \eta_k(x_i) \tilde{I}_k(x_i)\, ,
\end{align}
where $\tilde{I}_k(x_i)$ is the $k^{\mathrm{th}}$ realisation of the random intensity at pixel $x_i$ as measured by an ideal detector, and $\eta_k(x_i)$ is the $k^{\mathrm{th}}$ realisation of the noise at pixel $x_i$.  The expectation of $\mathfrak{G}^{(n)}(x_i)$ is then given by
\begin{align}\label{eq:mui}
\mu_i & = \braket{\mathfrak{G}^{(n)}(x_i, s_2 \dots,s_n)}  \\
& =  \frac{1}{N} \sum_{k=1}^{N} \braket{ I_k(x_1) I_k(s_2) \dots I_k(s_n) }  \nonumber \\
& =  \frac{1}{N} \sum_{k=1}^N \braket{ \tilde{I}_k(x_i) \tilde{I}_k(s_2)\dots \tilde{I}_k(s_n)} \nonumber \\
& \phantom{=} \times \braket{ \eta_k(x_i) \eta_k(s_2)\dots \eta_k(s_n)}\, , \nonumber
\end{align}
which is to be used in Eq.~(\ref{eq:loglike}). We assume that the noise and intensity are stationary random variables, so we can immediately perform the sum removing the factor $N^{-1}$.  The first factor in Eq.~\eqref{eq:mui} is, by definition, the intensity correlation $\braket{ \tilde{I}_k(x_i) \tilde{I}_k(s_2)\dots \tilde{I} _k(s_n)} = G^{(n)}(x_1, s_2, \dots,s_n)$.  The second factor in Eq.~\eqref{eq:mui} is in general some combination of moments of the normal distribution characterised by $\nu$ and $\varsigma$ (see Appendix).  

As shown in Fig.~\ref{fig:pixarray}a, we calculate the two-point intensity correlation $\frak{G}^{(2)}(x,s_2)$ between any pair of pixels $(x,s_2)$ as a function of the position of the pixel at $x$ and the stationary position of the second pixel at $s_2$.  Note, however, that when measuring a correlation function in this way the individual data points that are calculated may not be independent:  The correlation between any pair of data points, e.g., $\frak{G}^{(2)}(x_1,s_2)$ and $\frak{G}^{(2)}(x_2,s_2)$, depends on the correlations between all the measured intensities $I_k(x_1)$, $I_k(x_2)$ and $I_k(s_2)$; since the measurement relies on the statistical dependence of these intensities, the resulting values of $\frak{G}^{(2)}(x_1,s_2)$ and $\frak{G}^{(2)}(x_2,s_2)$ will not, in general, be statistically independent. The same argument holds for higher-order correlations.  In principle it is possible to avoid correlations between the data points altogether by taking each measurement in a completely independent manner.  Hanbury Brown and Twiss did just this in their original experiments. However, the price paid for taking data in such a way is a much greater total measurement time.  A more efficient way to collect the data would be to make use of an array of detectors (usually pixels of a CCD camera) to take all the measurements simultaneously.  As long as we are careful to take into account the correlations that arise in the data when measured in this manner then we are free to use this efficient method of data collection.  In the following we assume that the data is collected in such a way and are careful to calculate the correlations in the data explicitly.

The elements of the covariance matrix in Eq.~(\ref{eq:loglike}) are given by 
\begin{widetext}
\begin{align}\label{eq:Cij}
\mathbf{C}_{ij} & \equiv  \cov \left( \mathfrak{G}^{(n)}(x_i),\mathfrak{G}^{(n)}(x_j) \right) = \braket{\mathfrak{G}^{(n)}(x_i) \mathfrak{G}^{(n)}(x_j)} - \braket{\mathfrak{G}^{(n)}(x_i)} \braket{\mathfrak{G}^{(n)}(x_j)}  \nonumber \\
& = \frac{1}{N^2} \sum_{k,l=1}^N \braket{I_k(x_i) I_k(s_2) \dots I_k(s_n) I_l(x_j) I_l(s_2) \dots I_l(s_n)} - \mu_i \mu_j ,
\end{align}
\end{widetext}
where $\cov$ denotes the covariance between the measured intensity correlations at pixel $x_i$ and $x_j$. Since $k$ and $l$ label the individual images that are statistically independent, we can split the sum into two parts
\begin{widetext}
\begin{align}
\mathbf{C}_{ij} &  = \frac{1}{N^2} \Biggl( \sum_{k=1}^N \braket{I_k(x_i) I_k(s_2) \dots I_k(s_n) I_k(x_j) I_k(s_2) \dots I_k(s_n)} 
 + \sum_{\substack{k,l=1\\ k \neq l}}^N  \braket{I_k(x_i) I_k(s_2) \dots I_k(s_n)} \braket{I_l(x_j) I_l(s_2) \dots I_l(s_n)} \Biggr) \nonumber \\
 & \phantom{=}  - \mu_i \mu_j .
\end{align}
\end{widetext}
Since we treat the intensities as stationary random variables, we can simply perform the sums over $k$ and $l$ to obtain
\begin{align}\label{eq:covij}
\mathbf{C}_{ij} & =  \frac{1}{N} \bigg( \braket{ I(x_i) I(x_j) I(s_2)^{2} \dots I(s_n)^{2} } - \mu_i \mu_j \bigg) \\
& = \frac{1}{N} \left[G^{(2n)}(x_i,x_j,s_2,s_2,\dots,s_n,s_n)\right.  \nonumber \\
& \quad \times \left.\braket{\eta(x_i) \eta(x_j) \eta(s_2) \eta(s_2) \dots \eta(s_n ) \eta(s_n)} - \mu_i \mu_j \right] .\nonumber
\end{align}
We see that the covariances between our data $\mathfrak{G}^{(n)}(x_i)$ and $\mathfrak{G}^{(n)}(x_j)$ depend on correlation functions of order $2n$.  The term $\braket{\eta(x_i) \eta(x_j) \eta(s_2) \eta(s_2) \dots \eta(s_n ) \eta(s_n))}$ is evaluated in the Appendix.  We now have a complete characterisation of the probability distribution $p({\bf x}| \btheta)$ and can therefore calculate the Fisher information to determine the lower bound on the variance of our estimates of $\btheta$ via the CRB and can also perform a maximum likelihood estimation procedure to estimate the dimensions of the source. In the next section we present numerical simulations of this estimation procedure.

\section{Numerical Simulations}\noindent
To determine the performance of our estimates, we produce simulations of the experiment shown in Fig.~\ref{fig:geo}b, where a circular aperture of radius of $a = 100~\upmu$m emits uniform thermal light of wavelength $\lambda = 633~$nm.  The complex degree of coherence of such a source in the far field is given by Eq.~\eqref{eq:cdccirc}. To simulate the experiment we again make use of the optical equivalence theorem and the $P$-representation.  The $P$-representation for thermal light takes the form of a complex multivariate normal distribution.  The simulation of the experiment is then performed by sampling from a $2M$-dimensional normal distribution corresponding to the real and imaginary parts of $\alpha(x_1), \dots , \alpha(x_M)$, i.e., $\alpha(x)=a(x)+i b(x)$ with covariances
\begin{align}
\label{eq:covariances}
\braket{a(x_i) a(x_j)} & = \braket{b(x_i) b(x_j)} = \frac{1}{2} \braket{I}\; \gamma(x_i,x_j) \nonumber \\
\braket{a(x_i) b(x_j)} & = 0\, .
\end{align}
Here we have assumed a uniform far field intensity distribution of the thermal source, i.e., $\braket{I(x_i)} = \braket{I(x_j)} \equiv \braket{I}$. 

We include the effect of pixel noise by adding an additional normal random variable to each of the intensities with mean $\nu$ and variance $\varsigma^2$.  The intensity correlations are then calculated from the simulated field by means of Eq.~\eqref{eq:gnmeas}, which are in turn used to estimate the dimensions of the source parameters.  We require averaging over a large set of data in order to apply the central limit theorem and treat the data as normally distributed.  Here the parameter $\braket{I}$ is also treated as an unknown parameter to be estimated.  The importance of this cannot be overstated: if we instead treat $\braket{I}$ as a constant, any slight deviation from the exact value can lead to catastrophic failure of the estimation procedure. It is therefore imperative that the unknown parameter $\braket{I}$ should be treated as a nuisance parameter in order to perform the maximum likelihood estimation.  Although there appear to be four unknown parameters: $\bm{\theta} = (a,\braket{I},\nu,\varsigma)$, we will see in Section \ref{sec:constloss} and \ref{sec:randomloss} that we can often combine the parameters $\braket{I}$ and $\nu$ into the new parameter $\braket{I_{\rm eff}} =  \nu \braket{I}$, $\nu$ and $\varsigma$ can be combined into the new parameter $\chi=\nu/ \varsigma$.  These represent the effective intensity recorded by the detector in the presence of inefficiencies characterised by $\braket{\eta} =\nu$ and the ratio of the average pixel inefficiencies to the standard deviation of the inefficiencies.

\begin{figure}[t]
\centering
\includegraphics[width=8.5cm]{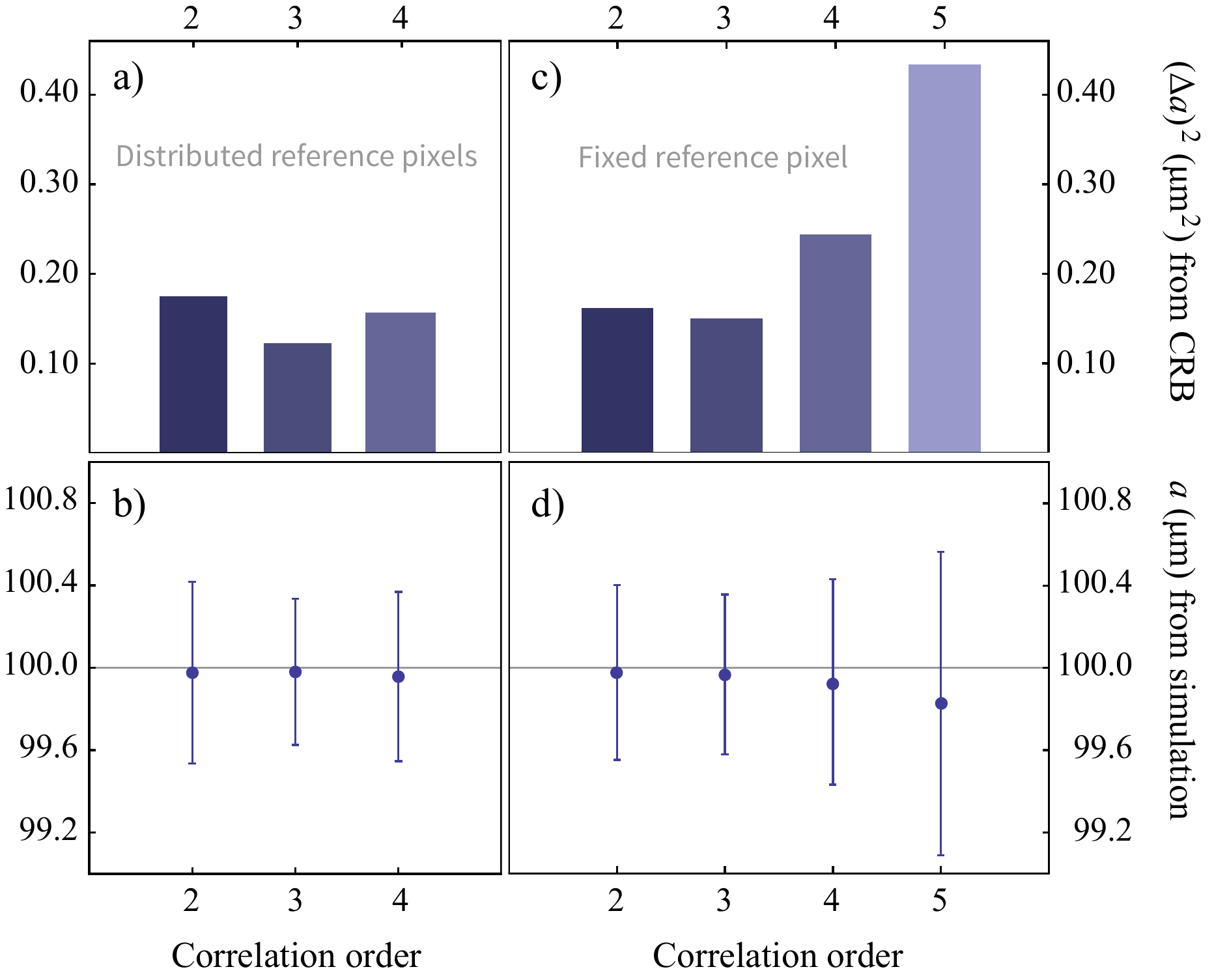}
\caption{(Color online) Results of the numerical estimation of the source diameter $a$ ($\upmu$m) and the corresponding variance $(\Delta a)^2$, in comparison with the Cram\'er-Rao bound (CRB). Choosing the reference pixels in a distributed manner as in Fig~\ref{fig:pixarray}c, e leads to a CRB value of the variance as shown in (a), with the estimate and standard deviation shown in (b). Here we have chosen a detection efficiency $\nu = 0.5$ with $\varsigma = 0.01$. Due to the computational complexity of the problem only correlation orders up to $n=4$ have been calculated. Choosing the central pixel as the reference pixel as in Fig~\ref{fig:pixarray}b, d leads to a CRB value of the variance as shown in (c), with the estimate and standard deviation shown in (d). This configuration does not allow us to include the detector efficiency as a random variable in the estimation procedure (see text for details). For comparison, the numerical values in this figure are collated in Tables~\ref{tab:var-nonoisemeas} and \ref{tab:var-noisemeas}.}
\label{fig:var-cl}
\end{figure}

Once we generated the simulated data from the $2M$-dimensional normal distribution described above and added the noise, we estimated the parameters $a$, $\braket{I_{\rm eff}}$ and $\chi$ based on different orders of intensity correlations and the scoring method.  The maximum likelihood estimation procedure was repeated 1000 times such that a statistical variance $(\Delta a)^2_{\rm Sim}$ for the simulated data could be calculated, and we can compare this to the lower bound on the variance $(\Delta a)^2_{\rm CRB}$ based on the Cram\'er-Rao bound:
\begin{align}
 (\Delta a)^2_{\rm CRB} \geq [ {\bf I }]^{-1}_{aa}\, ,
\end{align}
where we now must find the inverse of the $3 \times 3$ Fisher information matrix $\mathbf{I}$. There are two main cases to consider.

\subsection{Constant detector loss}\label{sec:constloss}\noindent
First, we analyse the special case of a detection system with constant loss for each pixel (i.e., $\varsigma =0$).  For this particular noise model, the choice of reference pixel positions does not affect the noise terms in Eqs.~(\ref{eq:mui}--\ref{eq:covij}), since $ \braket{\eta_k(x_i) \eta_k(s_2) \dots \eta_k(s_n)}= \nu^n$ for all choices of reference pixel positions.  We find the choice $s_2=s_3=\dots=s_n \equiv s = \lfloor M/2 \rfloor$ to be of particular interest since it simply involves taking powers of the measured intensity $I(s)^{n-1}$, and it is the central pixel on a one-dimensional CCD.  This allows us to compare the effects of the post processing without the need to consider complications regarding the exact placement of the reference pixels, $s_2, \dots, s_n$.  The inherent simplicity of this arrangement also allows us to calculate the correlations up to arbitrary order since the correlation functions $G^{(n)}$ and $G^{(2n)}$ take the compact forms
\begin{widetext}
\begin{align}
G^{(n)}(x_i,s,\dots,s) & = \braket{I}^n (n-1)! [1 + (n-1)|\gamma(x_i,s)|^2] , \\
G^{(2n)}(x_i,x_j,s,\dots,s) & =  \braket{I}^{2n} (n-2)! \big\{1 + |\gamma(x_i,x_j)|^2 \nonumber \\
& \phantom{=} + (n-2) \big[ 2 {\rm Re}\! \left( \gamma(x_i,x_j) \gamma(s,x_i) \gamma(x_j,s) \right) + |\gamma(x_i,s)|^2 +|\gamma(x_j,s)|^2  + |\gamma(x_i,s)|^2 |\gamma(x_j,s)|^2 \big] \big\} \nonumber .
\end{align}
\end{widetext}
We can therefore make the re-parameterisation suggested above, leaving us with the parameters $\btheta = (a,\braket{I_{{\rm eff}}}$) to estimate.  Fig.~\ref{fig:var-cl}c shows the Cram\'er-Rao lower bound on the variance for our estimate of $a$ for the four correlation functions $G^{(2)}$ to $ G^{(5)}$, and Fig.~\ref{fig:var-cl}d shows the estimate of $a$ with the actual standard deviation $\Delta a_{\rm Sim}$. The numerical results are also collated in Table~\ref{tab:var-nonoisemeas}.  
\begin{table}[b]
\caption{Results of the maximum likelihood estimation for the correlation functions $G^{(2)}$ to $G^{(5)}$ and the Cram\'er-Rao lower bounds.  The estimation procedure was performed on 1000 simulated data sets.\label{tab:var-nonoisemeas}}
\begin{ruledtabular}
\begin{tabular}{cccc}
 $n$ 	& $a$ $(\upmu {\rm m})$	& $(\Delta a)^2_{\rm Sim}~(\upmu {\rm m}^2)$	& $(\Delta a)^2_{\rm CRB}~(\upmu {\rm m}^2)$\medskip	\\ 
2      & 99.978	& 0.181 		& 0.162		\\
3	& 99.968	& 0.151 		& 0.150 		\\
4 	& 99.932	& 0.249		& 0.244 		\\
5	& 99.826	& 0.543 		& 0.434		
\end{tabular}
\end{ruledtabular}
\end{table}

\begin{figure}[t]
\centering
\includegraphics[width=8cm]{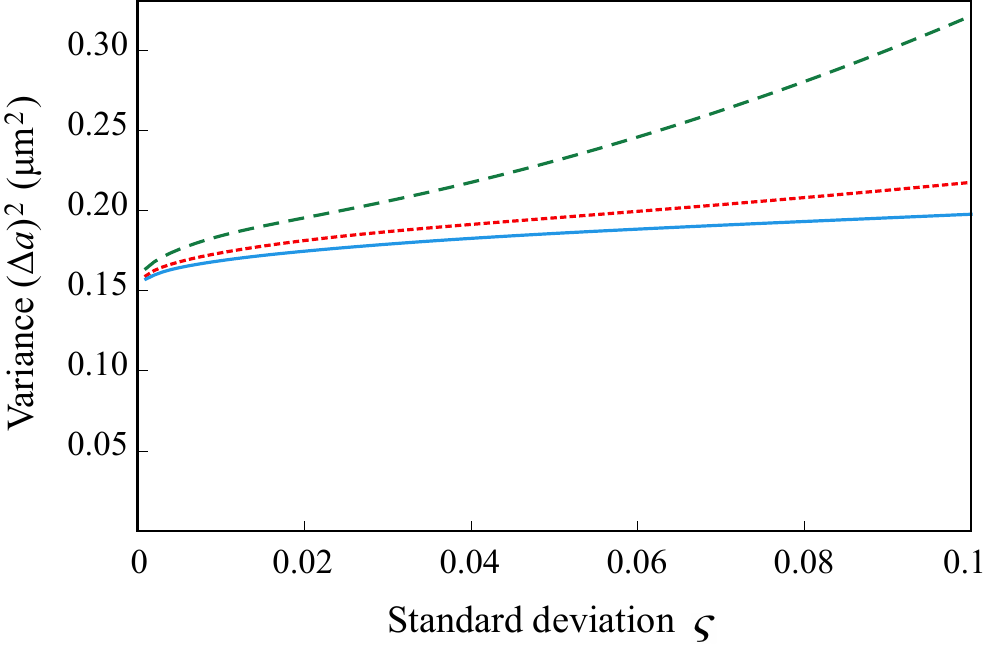}~
\caption{(Color online): Variance $(\Delta a)^2$ of the estimator $\hat{a}$ for $G^{(2)}$ against the standard deviation of the noise $\varsigma$ (dimensionless).  Average detector efficiencies $\nu = 0.2$ (\textcolor{OliveGreen}{dashed}), $\nu = 0.5$ (\textcolor{red}{dotted}), and $\nu = 0.9$ (\textcolor{MidnightBlue}{solid}).} \label{fig:fsigma}
\end{figure}

The best estimates of the spatial dimensions of the source occur for $n=3$, whereas the estimates get progressively worse as the correlation order is increased beyond third order.  Therefore, to extract the maximum amount of spatial information from our data, correlations of third order should be used.  We stress the importance of this finding as it requires no additional measurements to be made other than those made to measure $G^{(2)}$.  Indeed, in principle it would even be possible to use the data collected by Hanbury Brown and Twiss to measure the angular diameter of Sirius \cite{BrownTwiss56} with improved resolution.  We also note that there is nothing in our treatment that uniquely picks out the spatial correlation functions, in the same manner we could equally discuss temporal correlations (see, e.g., \cite{Monticone2014}).  Interestingly, estimates of the effective intensity $\braket{I_{\rm eff}}$ do not follow the same pattern as those for $a$.  If we wish to estimate $\braket{I_{\rm eff}}$ the best performance is given by $G^{(2)}$, with higher-orders performing worse.

\subsection{Detector loss as a random variable}\label{sec:randomloss}\noindent
Second, we demonstrate the effect of a small non-zero $\varsigma$, representing a system with uncertainty in the detector loss mechanism.  The effect of this additional noise is shown in Fig.~\ref{fig:fsigma}, where we plot the variance of the estimator for the second order intensity correlation function against the standard deviation $\varsigma$.  As expected, the addition of noise in the detection process reduces the precision in our estimator.

\begin{figure}
\centering
\includegraphics[width=8.5cm]{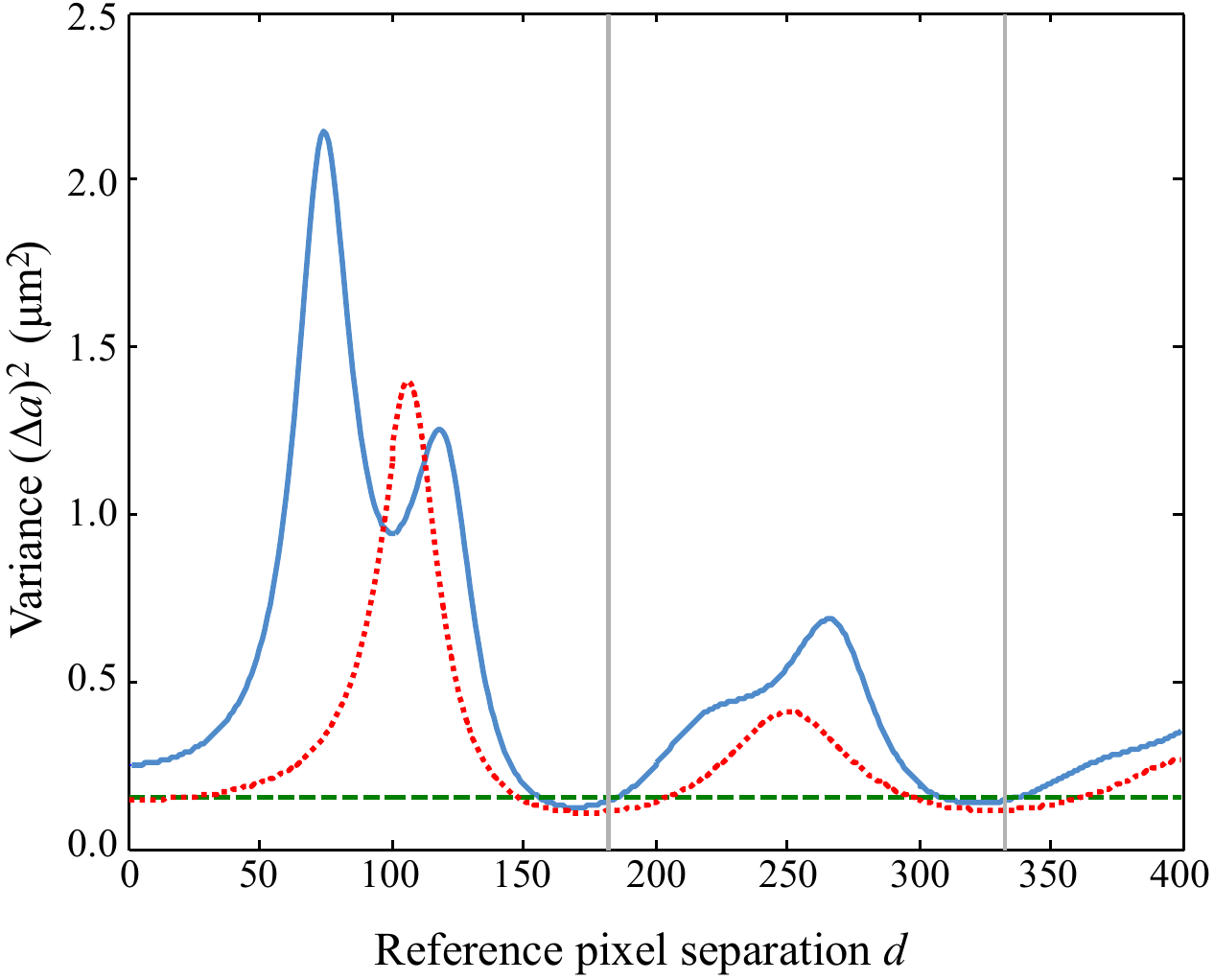}
\caption{(Color online): Standard deviation of the estimator $\hat{a}$ for $G^{(2)}$ (\textcolor{OliveGreen}{dashed}) , $G^{(3)}$ (\textcolor{red}{dotted}) and $G^{(4)}$ (\textcolor{MidnightBlue}{solid}) as a function of  reference pixel separation $d=| s_i - s_{i+1}|$ for detection scheme~2, given a circular aperture of radius $a = 100~\upmu$m.  The grey vertical lines correspond to the first and second zeroes of the Bessel function $J_1$. } \label{fig:SD-bessel}
\end{figure}

In order to perform the estimation, first we must evaluate the second term in Eq.~\eqref{eq:mui}, which is the $n^{{\rm th}}$ moment of the noise distribution.  Having considered the case where all reference pixels are the same in the previous section, we now restrict ourselves to only considering cases where no two reference pixel positions are the same, i.e. $s_2 \neq s_3 \neq \dots \neq s_n$, such that we can determine the effect of separating the reference pixels.  In this regime the $n^{{\rm th}}$ moment of the noise distribution, $\braket{\eta(x_i) \eta(s_2) \dots \eta(s_n)}$, is given by
\begin{align}\label{eq:expnoise}
\braket{\eta_k(x_i)\eta_k(s_2) \dots \eta_k(s_n)}& = \nu^n + \nu^{n-2}\sum_{j=2}^n \delta_{x_i s_j} \varsigma^2 \nonumber \\
& = \nu^n\left(1 + \sum_{j=2}^n \delta_{x_i s_j} \chi^2 \right).
\end{align}
We therefore find it necessary to re-parameterise the problem using the parameters $\btheta = (a,\braket{I_{{\rm eff}}},\chi)$ as mentioned above.  Eq.~\eqref{eq:expnoise} represents an $n^{\rm th}$ moment of the noise distribution, and the Kronecker deltas arise from the independence of the distribution for individual pixels.  The $2n^{{\rm th}}$ moment of the noise distribution appearing in Eq.~\eqref{eq:covij} is calculated in the Appendix.

\begin{figure}
\centering
\includegraphics[width=8.5cm]{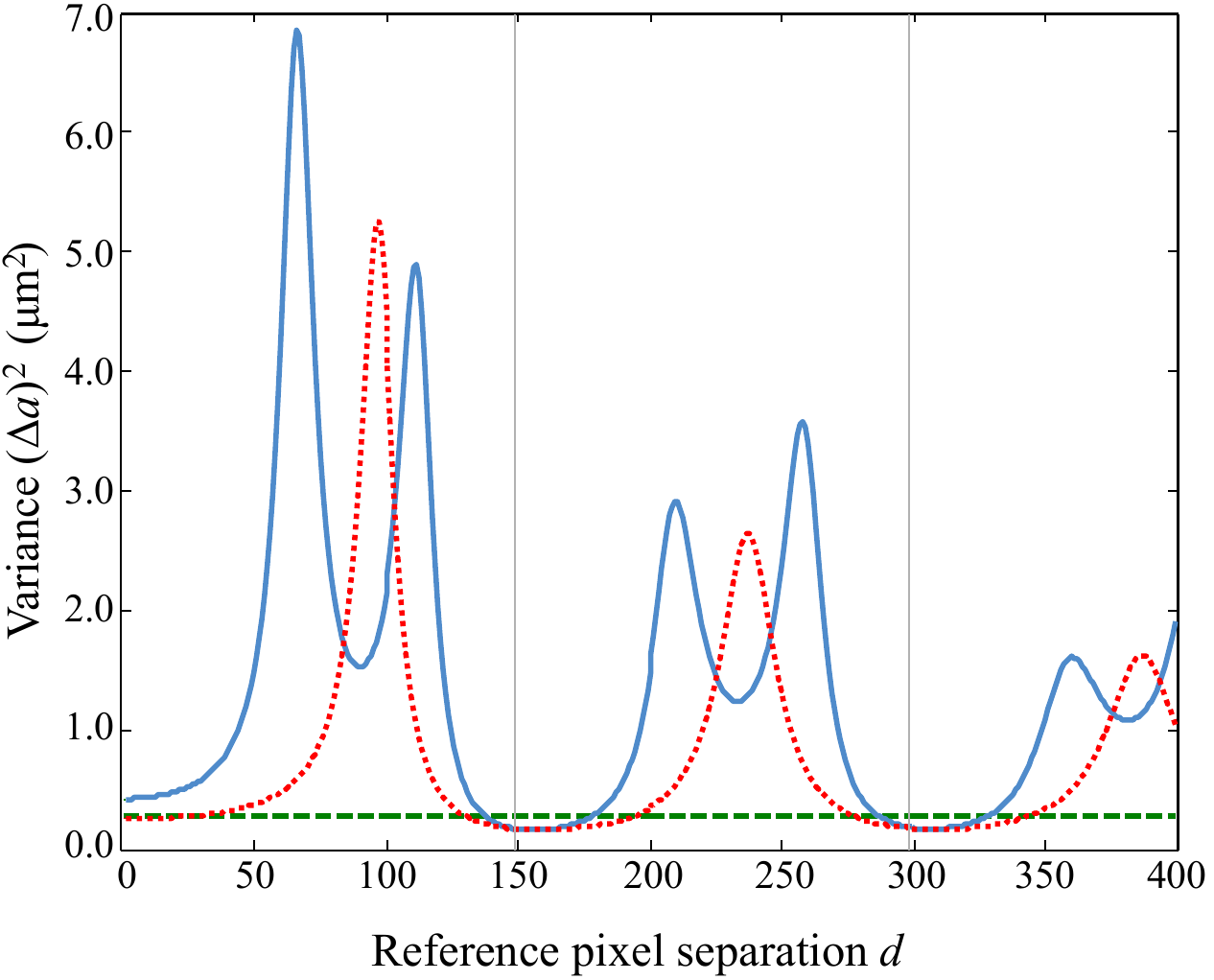}
\caption{(Color online): Standard deviation of the estimator $\hat{a}$ for $G^{(2)}$ (\textcolor{OliveGreen}{dashed}) , $G^{(3)}$ (\textcolor{red}{dotted}) and $G^{(4)}$ (\textcolor{MidnightBlue}{solid})  as a function of  reference pixel separation $d = | s_i - s_{i+1}|$ for detection scheme~2, given a slit of width $a = 200~\upmu$m.  The grey vertical lines correspond to the first and second zeroes of the ${\rm sinc}$ function.}\label{fig:varsinc}
\end{figure}

\begin{table}[b]
\caption{Results of the maximum likelihood estimation for the correlation functions $G^{(2)}$ to $G^{(4)}$ and the Cram\'er-Rao lower bounds, with a reference pixel separation of $d=182$ that corresponds to the first zero of $J_1$.  The estimation procedure was performed on 1000 simulated data sets.\label{tab:var-noisemeas}}
\begin{ruledtabular}
\begin{tabular}{cccc}
 $n$ 	& $a$ $(\upmu {\rm m})$	& $(\Delta a)^2_{\rm Sim}~(\upmu {\rm m}^2)$	& $(\Delta a)^2_{\rm CRB}~(\upmu {\rm m}^2)$\medskip	\\
2      & 99.976  	& 0.194 		& 0.175 		\\
3	& 99.980	& 0.126 		& 0.123 		\\
4 	& 99.957	& 0.169		& 0.157 		
\end{tabular}
\end{ruledtabular}
\end{table}

In order to find the optimum position of the reference pixels, we define the dimensionless number $d = | s_i - s_{i+1}|$, the separation between adjacent reference pixels, and plot the standard deviation as a function of $d$.  Fig.~\ref{fig:SD-bessel} shows the standard deviation for $G^{(2)}$ to $G^{(4)}$ as a function of $d$.  Interestingly, the higher-order correlations outperform $G^{(2)}$ only for some values of $d$.  For $G^{(3)}$ we find that the optimum positions correspond to separations where the two reference pixels become uncorrelated.  This occurs whenever the complex degree of coherence between the two pixels is approximately equal to zero.  Since the complex degree of coherence for the system is proportional to the Bessel function $J_1$, the optimum separations $d$ correspond to the zeroes of this function.  For $G^{(4)}$ the exact position of the optimum is more complicated, due to the fact that the zeroes of $J_1$ are not uniformly distributed.  However, the optimum positions are approximately located at the position where adjacent reference pixels are uncorrelated from their nearest neighbours. Tab.~\ref{tab:var-noisemeas} shows the variance of the estimators for the first three correlation functions as calculated from the CRB and directly measured in the simulations.  We see that the measured variance in our estimators closely follows that obtained from the CRB.   The results of the maximum likelihood estimation and the Cram\'er-Rao bound are given in Table~\ref{tab:var-noisemeas}.

Another interesting feature of Fig.~\ref{fig:SD-bessel} is the ability for $G^{(4)}$ to outperform $G^{(2)}$, a feature that does not occur for fixed reference pixels. This behaviour is reminiscent of the ``magic angles'' in Refs.~\cite{Thiel07,Oppel12}, where the detectors had to be placed at different specific positions (the magic angles) in order to obtain the $(n-1)$-fold increased sinusoidal modulation in the scanning detector.

The exact relation between the variance of the estimators and the correlation order also depends on the geometry of the source.  Fig.~\ref{fig:varsinc} shows the dependence of the variance as a function of the reference pixel separation for a \emph{slit} of width $a=200~\upmu$m.  The complex degree of coherence for such a geometry is given by the ${\rm sinc}$ function. Since the zeros of the ${\rm sinc}$ function are uniformly distributed, it is possible to achieve independence for all the reference pixels simultaneously.  Fig.~\ref{fig:varsinc} shows that for the optimal choice of $d$ the estimator for $G^{(4)}$ outperforms $G^{(2)}$ and is about as good as $G^{(3)}$.

\section{Discussion \& conclusions}\noindent
We have discussed the exact role of higher-order intensity correlations with respect to parameter estimation of the intensity distribution of thermal sources, and demonstrated that it is beneficial to post-process the data in such a way as to measure intensity correlations of order $n>2$.  We have also shown how the post processing can be optimised with respect to the placement of the reference pixels, in order to find the most informative measurements.  A major benefit of this method is that it does not require particularly elaborate experimental arrangements.  Indeed, in certain circumstances it is even possible to increase the precision simply by taking powers of the measured intensities.  Since we explicitly account for correlations between the data points [see Eqs.~(\ref{eq:Cij}-\ref{eq:covij})], all the measurements can be made simultaneously, thus reducing considerably the measurement time required to obtain the data. While we have framed the discussion in the context of detector pixels on a CCD camera, the same methods apply to any array of field detectors, including telescopes. By fully determining the probability distribution function (PDF) for measurements of intensity correlation functions, including the covariance matrix of the correlated data, we are able to determine the Fisher information for such experiments.  This allows us to calculate the maximum achievable precision via the Cram\'er-Rao bound and also to saturate that bound by performing a maximum likelihood estimation.  

The techniques presented here can in principle be used to estimate the source dimensions of any object that emits incoherent light, including quantum emitters. In many cases the Gaussian moment theorem does not apply and extra care must be taken in the calculation of the correlation functions $G^{(n)}(x_1,s_2,\ldots,s_n)$.  As long as the measured intensity is a random variable then the PDF for the data can be considered again as a multivariate normal when averaged over many measurements.  Also, the estimated parameters need not be spatial parameters of the source: as long as the PDF depends on the parameters to be estimated in a deterministic way our procedure can be used to perform the estimation.  The inclusion of a realistic noise model allows the effect of detection noise to be accounted for, thus allowing the estimation to proceed in the presence of noise.

The experimental implementation of this procedure is not exceedingly challenging, as it requires the same kind of techniques as used in the original HBT experiment some 60 years ago. A number of critical points must be met though.  First, when measuring the intensity correlation functions, the integration time of the detectors must be well below the coherence time of the radiation to ensure that each measurement captures a single longitudinal mode of the radiation. In addition, the area of each detector (pixel) must be much smaller than the coherence area of the source such that a single pixel can be considered as measuring a single transversal mode. This ensures that every detection event samples no more than a single mode volume (i.e., the speckle size). Second, our use of the central limit theorem implies that we have to use a large data set that reduces statistical uncertainties to a minimum and provides a good signal to noise ratio. This is a well-known requirement for any maximum likelihood estimation procedure. Third, we have assumed uniform mean intensity across the entire CCD. In practice this can be challenging as a natural source might have a non-uniform intensity profile, or there might be additional spurious interference effects at the detector (e.g., an etalon effect due to the protective glass of a CCD camera).
Finally, we have assumed that the detector noise is uncorrelated, which means that the pixel efficiencies can be considered random and there is no cross-talk between the pixels. The latter two requirements can in principle be included in the modelling of the experiment, but this comes at the cost of a significantly increased complexity of the correlation functions.

The ability of higher-order intensity correlations to display more information than lower order correlations is often a source of confusion.  In fact, the original Hanbury Brown and Twiss experiment caused great controversy, and its eventual resolution heralded the beginning of quantum optics as a mature discipline \cite{silva13}. The improved estimation capability is most clearly demonstrated when the reference pixels are all the same $s_2 = s_3 = \dots = s_n$.  We can then take the output of two photodetectors and simply by taking powers of one of the outputs we can gain a more precise estimate of the angular diameter of the source.  We could understand this increase in precision by comparing the measurements of the first and second order correlation function as in the original HBT experiment.  Here we use the same set of data when measuring the first order correlation function, i.e., the intensity distribution in the far field, as we use to measure the second order intensity correlation, and yet a measurement of the intensity reveals almost no information about the source since the intensity of a thermal source in the far field is constant across $x$.  In contrast, the second order intensity correlation function is highly dependent on $x$, which allows for a lensless measurement and therefore a precise estimate of the angular diameter of the source (see Fig.~\ref{fig:G2plot}).  When considered as another method of post processing the data, it is no more surprising that higher-order correlations outperform the second order correlation than the second order correlation outperforming first-order intensity measurements.

Our method is rather general, and there is nothing in our treatment that uniquely picks out the spatial correlation functions, in the same manner we could equally discuss temporal correlations. Future work will focus on (multi-mode) squeezed light and single photon sources. Since all correlation functions can be determined from the same data set, it would be advantageous to combine all of the estimates achieved via different $n$ into a single estimate. In order to do this properly we would need to know exactly how all of the individual estimates are correlated to determine the appropriate weights for the combined estimate and its error.  The difficulty in determining the correlation is in knowing how the maximum likelihood estimation procedure affects the correlation, if at all. Once this is known, it should be possible to determine the weights and obtain the final estimate. 

In conclusion, we found that higher-order correlation functions can substantially improve estimation of the parameters that characterise the geometry of an incoherent light source. As long as the source and the detection system are properly modelled, the procedure can be implemented with current technology. 

\section{Acknowledgements}\noindent
The authors would like to thank Earl Campbell, Sam Coveney, Michael Woodhouse, Tom Bullock, Alex Safar and Steven M. Kay for fruitful discussions and advice.

\begin{appendix}

\section{Moments of the noise distribution}\noindent
In this Appendix we evaluate the moments of the noise distribution for the general case and for the case of a Gaussian noise distribution as considered in Section IV.  First, we need to evaluate the term $\braket{\eta(x_i) \eta(s_2) \dots \eta(s_n)}$ which appears in Eq.~\eqref{eq:mui}.  In Section V we calculated this term for the Gaussian noise distribution, here we calculate it for the general case.  Denoting $J_1 = \braket{\eta(s_2)} \dots \braket{\eta(s_n)}$ and $S = \{ s_2, \dots, s_n\} $ we find
\begin{align}
\braket{\eta(x_i) \eta(s_2) \dots \eta(s_n)} = \begin{cases} \braket{\eta(x_i)} J_1 & {\rm if} \,\, x_i \notin S \\
\frac{\braket{\eta(x_i)^2}}{\braket{\eta(x_i)}} J_1 & {\rm if} \,\, x_i \in S
\end{cases},
\end{align}
if none of the reference pixels are equal, i.e., $s_2 \neq s_3 \neq \dots \neq s_n$.  If instead we use detection scheme 1, where all the reference pixels are the same, we find
\begin{align}
\braket{\eta(x_i) \eta(s_2)^{n-1}} = \begin{cases} \braket{\eta(x_i)} J_2 & {\rm if} \,\, x_i \neq s_2 \\
\braket{\eta(x_i)^{n}}  & {\rm if} \,\, x_i = s_2
\end{cases},
\end{align}
where $J_2 = \braket{\eta(s_2)^{n-1}}$. 

Now we evaluate the second term in Eq.~\eqref{eq:covij}, $\braket{ \eta(x_i) \eta(x_j) \eta(s_2)^{2} \dots \eta(s_n)^{2}}$.  Since the noise is treated as uncorrelated between the pixels, the expectation value factorises into $\braket{\eta(x_i)} \braket{ \eta(x_j)} \braket{\eta(s_2)^{2}} \dots \braket{ \eta(s_n)^{2}}$ if $x_i$ and $x_j$ are not equal to each other or any of the positions $s_2, \dots , s_n$.  However, more generally, the expression is more complicated.  Denoting $J_3 = \braket{\eta(s_2)^2} \dots \braket{\eta(s_n)^2}$ we obtain
\begin{widetext}
\begin{align}\label{eq:2nmom}
\braket{ \eta(x_i) \eta(x_j) \eta(s_2)^{2} \dots \eta(s_n)^{2}} = \begin{cases} \frac{\braket{\eta(x_i)^4}}{ \braket{\eta(x_i)^2}} J_3 = a  & {\rm if } \,\, x_i=x_j \in S \\
 \braket{\eta(x_i)^2} J_3 = b &  {\rm if } \,\, x_i=x_j \notin S  \\
  \frac{\braket{\eta(x_i)^3}}{ \braket{\eta(x_i)^2}} \frac{\braket{\eta(x_j)^3}}{ \braket{\eta(x_j)^2}} J_3 = c &  {\rm if } \,\, x_i \neq x_j ; \,\, x_i,x_j \in S  \\
 \braket{\eta(x_i)}  \braket{\eta(x_j)} J_3 = d & {\rm if } \,\,  x_i \neq x_j ; \,\, x_i,x_j  \notin S  \\
 \frac{\braket{\eta(x_i)^3}}{ \braket{\eta(x_i)^2}} \braket{\eta(x_j)} J_3 = e &  {\rm if } \,\,  x_i \neq x_j ; \,\, x_i \in S ; \,\, x_j \notin S  \\
\braket{\eta(x_i)} \frac{\braket{\eta(x_j)^3}}{ \braket{\eta(x_j)^2}} J_3  = f & {\rm if } \,\,  x_i \neq x_j ; \,\, x_i \notin S ; \,\, x_j \in S 
\end{cases},
\end{align}
and for detection scheme 1
\begin{align}
\braket{ \eta(x_i) \eta(x_j) \eta(s_2)^{2n-2}} = \begin{cases} \braket{\eta(x_i)^4}  & {\rm if } \,\, x_i=x_j = s_2  \\
 \braket{\eta(x_i)^2} J_4  &  {\rm if } \,\, x_i=x_j \neq s_2  \\
 \braket{\eta(x_i)} \braket{\eta(x_j)} J_4   & {\rm if } \,\,  x_i \neq x_j ; \,\, x_i \neq s_2 ; \,\, x_j \neq s_2 \\
 \braket{\eta(x_i)} \braket{\eta(x_j)^{2n-1}}  &  {\rm if } \,\,  x_i \neq x_j = s_2 \\
 \braket{\eta(x_j)} \braket{\eta(x_i)^{2n-1}}  &  {\rm if } \,\,  x_j \neq x_i = s_2 
\end{cases},
\end{align}
\end{widetext}
where $J_4 = \braket{\eta(s_2)^{2n-2}}$.  For a general noise distribution, each term in these piecewise functions can be associated with a parameter to be estimated.  The use of a general  noise distribution can therefore be incorporated in the theory but comes at the expense of a greater number of estimation parameters.  The benefit of using a Gaussian noise model is that there are only two additional noise parameters corresponding to the first and second moments of the Gaussian distribution, or a combination of them as in section V, ($\nu$, $\chi$).

To give a more visual presentation of the noise correlations, we can represent the resulting piecewise function, Eq. (A.3), in matrix form ${\bf M}_{ij}=\braket{ \eta(x_i) \eta(x_j) \eta(s_2)^{2} \dots \eta(s_n)^{2}}$ 
\begin{align}
{\bf M} =
\begin{pmatrix}
 {\textcolor{red} b} & {\textcolor{OliveGreen} d} & {\textcolor{OliveGreen} d} & e & {\textcolor{OliveGreen} d} & {\textcolor{OliveGreen} d} & {\textcolor{OliveGreen} d} & e & \dots \\
 {\textcolor{OliveGreen} d} & {\textcolor{red} b} & {\textcolor{OliveGreen} d} & e & {\textcolor{OliveGreen} d} & {\textcolor{OliveGreen} d} & {\textcolor{OliveGreen} d} & e & \dots \\
 {\textcolor{OliveGreen} d} & {\textcolor{OliveGreen} d} & {\textcolor{red} b} & e & {\textcolor{OliveGreen} d} & {\textcolor{OliveGreen} d} & {\textcolor{OliveGreen} d} & e & \dots \\
 e & e & e & {\textcolor{MidnightBlue} a} & e & e & e & {\textcolor{orange} c} & \dots \\
 {\textcolor{OliveGreen} d} & {\textcolor{OliveGreen} d} & {\textcolor{OliveGreen} d} & e & {\textcolor{red} b} & {\textcolor{OliveGreen} d} & {\textcolor{OliveGreen} d} & e & \dots \\
 {\textcolor{OliveGreen} d} & {\textcolor{OliveGreen} d} & {\textcolor{OliveGreen} d} & e & {\textcolor{OliveGreen} d} & {\textcolor{red} b} & {\textcolor{OliveGreen} d} & e & \dots \\
{\textcolor{OliveGreen} d} & {\textcolor{OliveGreen} d} & {\textcolor{OliveGreen} d} & e &{\textcolor{OliveGreen} d} & {\textcolor{OliveGreen} d} &  {\textcolor{red} b} & e & \dots \\
 e & e & e & {\textcolor{orange} c} & e & e & e & {\textcolor{MidnightBlue} a} & \dots \\
 \vdots & \vdots & \vdots & \vdots & \vdots & \vdots & \vdots & \vdots & \ddots \\
\end{pmatrix},
\end{align}
where we have made use of the fact that $f=e$ since the noise is assumed to be the same for all pixels.

\end{appendix}

\bibliographystyle{apsrev4-1}

\end{document}